\documentclass[12pt,preprint]{aastex} % Submission
\shorttitle{Koenigstuhl 1, 2, and 3}
\shortauthors{J. A. Caballero}

\begin{document}
\title{Southern wide very low-mass stars and brown dwarfs in resolved binary and
multiple systems}

\author{Jos\'e Antonio Caballero\altaffilmark{1}}
\affil{Max-Planck-Institut f\"ur Astronomie, K\"onigstuhl 17, D-69117
Heidelberg, Germany}
\email{caballero@mpia.de}

\altaffiltext{1}{Alexander von Humboldt Fellow at the MPIA.}

\begin{abstract}
The results of the Koenigstuhl survey in the Southern Hemisphere are presented. 
I have searched for common-proper motion companions to 173 field very
low-mass stars and brown dwarfs with spectral types $>$ M5.0V and magnitudes $J
\lesssim$ 14.5\,mag.  
I have measured for the first time the common-proper motion of two new wide
systems containing very low-mass components, Koenigstuhl~2~AB and
3~A--BC.  
Together with Koenigstuhl~1~AB and 2M0126--50AB, they are among the widest
systems in their respective classes ($r$ = 450--11\,900\,AU). 
I have determined the minimum frequency of field wide multiples ($r >$ 100\,AU)
with late-type components at 5.0$\pm$1.8\,\% and the frequency of
field wide late-type binaries with mass ratios $q >$ 0.5 at
1.2$\pm$0.9\,\%. 
These values represent a key diagnostic of evolution history and low-mass star
and brown-dwarf formation scenarios. 
 Additionally, the proper motions of 76 field very low-mass dwarfs
are measured here for the first time.
\end{abstract}


\section{Introduction}
\label{intro}

Very low-mass (VLM) dwarfs have  masses of about one tenth of the
Solar mass or less, and spectral types later than M5 V ($T_{\rm eff}
\lesssim$ 3000\,K). 
 Many of them are found in binary and multiple systems with a large variety
of separations and mass ratios.   
Proxima Centauri (M5.5V), with a mass of 0.11$\pm$0.02\,$M_\odot$, is the
nearest and most famous example of a VLM dwarf in a multiple
system.  
Given its large separation to \object{$\alpha$ Cen}~AB, of more than
15\,000\,AU, Proxima is close to be gravitationally unbound (Wertheimer \&
Laughlin 2006 and references therein). 
The binary BL~Ceti + UV~Ceti (M5.5V+M6.0V), the sixth closest star system to the
Sun, is on the contrary a tight binary separated by only $\sim$5\,AU (Heintz 
1987). 
There are even some VLM field dwarfs that are both tight binaries and
companions to more massive stars, e.g. \object{$\epsilon$~Indi~BC} 
(T1V+T6V)  at $\sim$1500\,AU to the nearby K4.5V star
\object{$\epsilon$~Indi~A (Scholz et al. 2003;} McCaughrean et al.~2004).  

 The systems containing VLM components can be dichotomized into two groups
according to their mass ratios.
One group comprises systems with mass ratios $M_2/M_1 \equiv q < 0.5$, and
includes from radial-velocity, transit and microlensing exoplanet candidates to
late-type star and brown-dwarf companions to FGKM-type stars detected in direct
image (see The Extrasolar Planets Encyclopaedia and Burgasser, Kirkpatrick \&
Lowrance 2005 for comprehensive compilations of planetary and late-M-, L-,
and T-type companions to stars, respectively). 
The other group contains late-type stars and brown dwarfs in double systems 
with mass ratios $q > 0.5$.
Throughout this work, I will refer to them as equal-mass VLM binaries (or simply
VLM binaries; see the Very Low Mass Binaries Archive maintained by Nick Siegler
containing an up-to-date list of stellar and substellar binary systems with
estimated total masses $M_1 + M_2 <$ 0.2\,$M_\odot$). 

The vast majority of the equal-mass VLM binaries yet found have relatively
small angular separations (of less than 1\,arcsec) and can be only
resolved with the {\em Hubble} Space Telescope or Adaptive Optics systems (e.g.
Bouy et al. 2005; Siegler et al. 2005).   
If the \object{SE~70} + \object{S\,Ori~68} system  in the $\sigma$~Orionis
cluster is not considered (without proper-motion confirmation; Caballero et al.
2006), there are only six known VLM binaries with $M_1 + M_2 <$
0.2\,$M_\odot$ separated by more than 50\,AU.   
Three are in very young star-forming regions (Ophiuchus, Chamaeleon~I), 
which probably will not survive the tidal disruption field within the clusters,
and three are field  VLM binaries.
The latters are \object{DENIS--P J055146.0--443412} ( DE0551--44AB;
$r \approx$ 220\,AU; Bill\`eres et al. 2005), \object{Koenigstuhl~1}~AB
(K\"o~1AB; $r =$ 1800\,$\pm$\,170\,AU; Caballero 2007) and \object{2MASS
J012655.49--502238.8} + \object{2MASS J012702.83--502221.1} (2M0126--50AB; $r =$
5100\,$\pm$\,400\,AU; Artigau et al. 2007).  
%$q \approx$ 0.93, $q =$ 0.77 $\pm$ 0.06, $q =$ 0.97$^{+0.03}_{-0.07}$
 Their mass ratios and total masses vary in the intervals 0.77 $\lesssim q
\lesssim$ 0.97 and 0.17\,$M_\odot \lesssim M_1 + M_2 \lesssim$ 0.19\,$M_\odot$,
respectively.  
There are other known VLM multiple systems in the field with
separations larger than  1000\,AU.
However, their total masses are several times larger than those of the
equal-mass VLM binaries and their mass ratios $q$ significatively deviate
from unity. 
 For example, the mass ratio between \object{vB~8} (M7.0V, $r
\sim$ 1\,400\,AU) and \object{GJ~644}~A--BD + \object{GJ~643} in the
\object{V1054~Oph} quintuple system is $q \approx 0.065$, and its total mass is
about 1.3\,$M_\odot$ (Kuiper 1934; Weis 1982; D'Antona 1986; S\"oderhjelm 1999;
Mazeh et al.~2001).   

2M0126--50AB and K\"o~1AB, whose secondary has a mass at the substellar
boundary, are by far the widest equal-mass VLM binaries yet found in the
field and are part of a new differentiated binary class.
Their separations are orders of magnitude larger than those of the VLM
tight binaries. 
They represent a challenge for the widely accepted idea that lighter
systems tend to have smaller separations (Sterzik \& Durisen 2004) and for
the ``embryo-ejection'' scenario of formation of substellar objects (Reipurth \&
Clarke 2001; Bate, Bonnell \& Bromm 2003). 
Large hydrodynamical simulations can produce wide low-mass binary systems,
albeit rarely.
Bate \& Bonnell (2005) showed an exotic situation in which two low-mass M dwarfs
(about 0.18\,$M_\odot$ each) were almost simultaneously ejected with similar
velocities from a small group of protostars.
As the two objects moved away from the group, it turned out that they were
weakly bound into a wide binary system. 
Their binding energy was, however, $\sim$4.4 times larger than that of
K\"o~1AB and $\sim$12 than that of 2M0126--50AB.
Further discussion on how wide equal-mass VLM binaries represent a key
diagnostic of star formation theories can be found in Bill\`eres et al. (2005),
Phan-Bao et al. (2005), Burgasser et al. (2007), Caballero (2007),
and Artigau et al. (2007). 
Caballero (2007) suggested that the wide separation between the components of
K\"o~1AB might be also due to perturbation resulting from encounters with
more massive objects as they traveled in the Galaxy, and not only to the
formation mechanism.
 
Both 2M0126--50AB and K\"o~1AB are exceptional binaries, but it is not
known yet {\em how} rare they are. 
In this  work, I present the results of the Koenigstuhl survey of wide
 VLM dwarf binaries in the Southern Hemisphere and measure
for the first time the frequency of equal-mass VLM systems.
It complements the near-infrared proper motion search for companions to
K5.0V--M7.0V stars at separations $\sim$100--1400\,AU carried out by Hinz et al.
(2002) and the Cerro Tololo Inter-American Observatory Parallax
Investigation of nearby multiples, primarily M dwarfs, by Jao et al.~(2003).

\section{The Koenigstuhl survey}

I have performed a proper-motion survey, called Koenigstuhl, using the the
UKST and POSS--I plates and the SuperCOSMOS Science Archive (SSA; Hambly et al.
2001a).  
The survey is limited to declination $<$ +3\,deg, where SSA data are only
available. 
I have investigated 173 VLM field dwarfs with spectral types between
M5.5V and L8.0V and brighter than $J \sim$ 14.5\,mag.
Their names, coordinates, proper motions, and spectral types are provided in
Table~\ref{investigated} (if tight binaries, only one spectral type is given).
The bulk of them were taken from Cruz et al. (2003) and Phan-Bao \& Bessel
(2006).
The SSA proper-motion measurements are accurate to $\sim$10\,mas\,a$^{-1}$ at
photograhic $B_J$, $R \sim$ 19--18\,mag and to $\sim$50\,mas\,a$^{-1}$ at $B_J$,
$R \sim$ 22--21\,mag (Hambly et al. 2001b), which are the expected magnitudes of
the faintest investigated VLM dwarfs.
Three nearby stars are however too bright in the SuperCOSMOS images (Proxima
Centauri, BL Cet + UV Cet, and EZ~Aqr, which saturate in the digitized
photographic plates and whose proper motions are not tabulated by SuperCOSMOS).
I have taken the values of their proper motions from Perryman et al. (1997) and
Salim \& Gould (2003). 
Besides, I have not identified in the SSA data \object{L~143--23} (M5.5V; with a
low galactic latitude), \object{HD~16270~B} (L1.0V; in the glare of the K3.0V
primary), and several mid-L dwarfs fainter than $J \sim$ 13.0\,mag. 

\begin{figure}
\plottwo{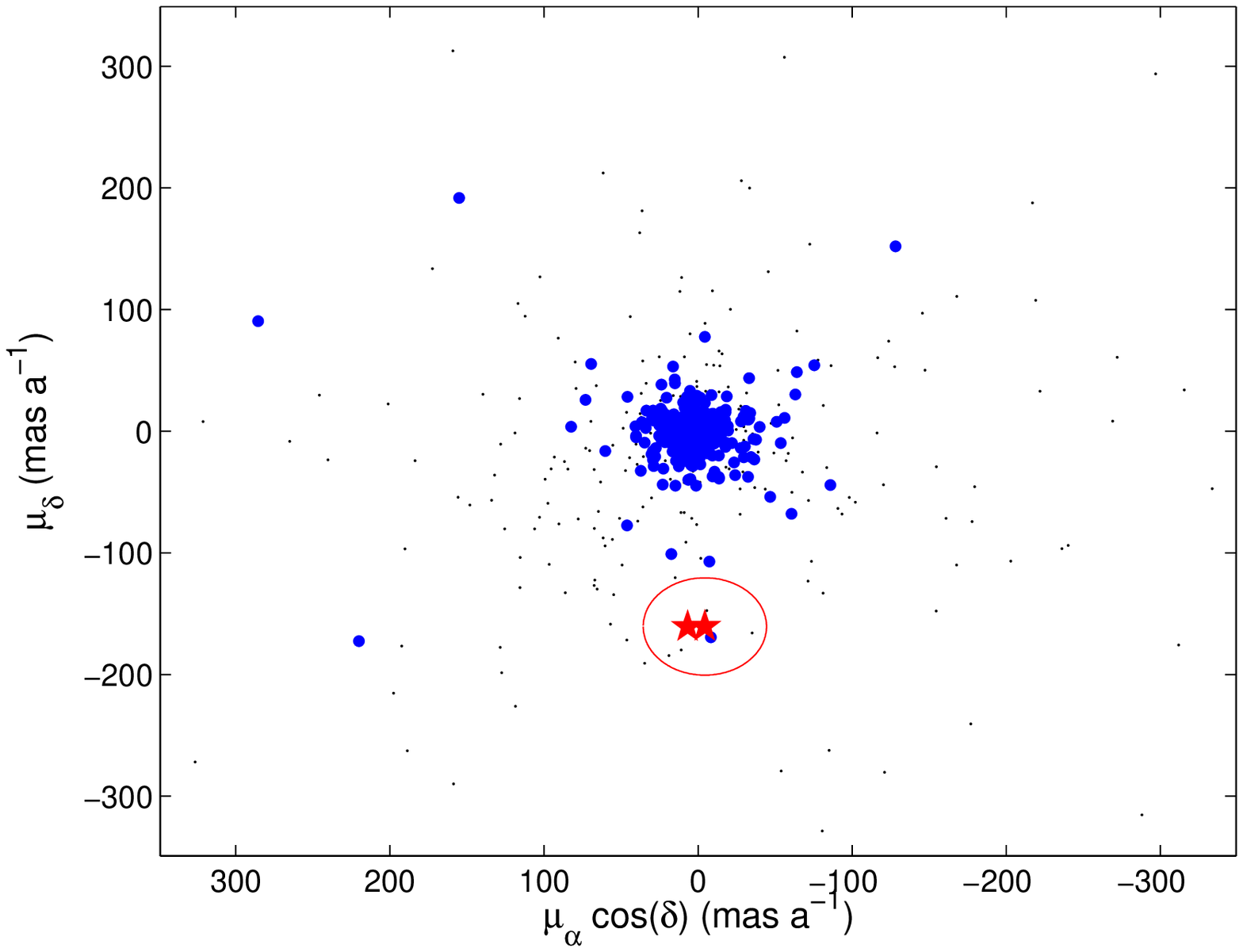}{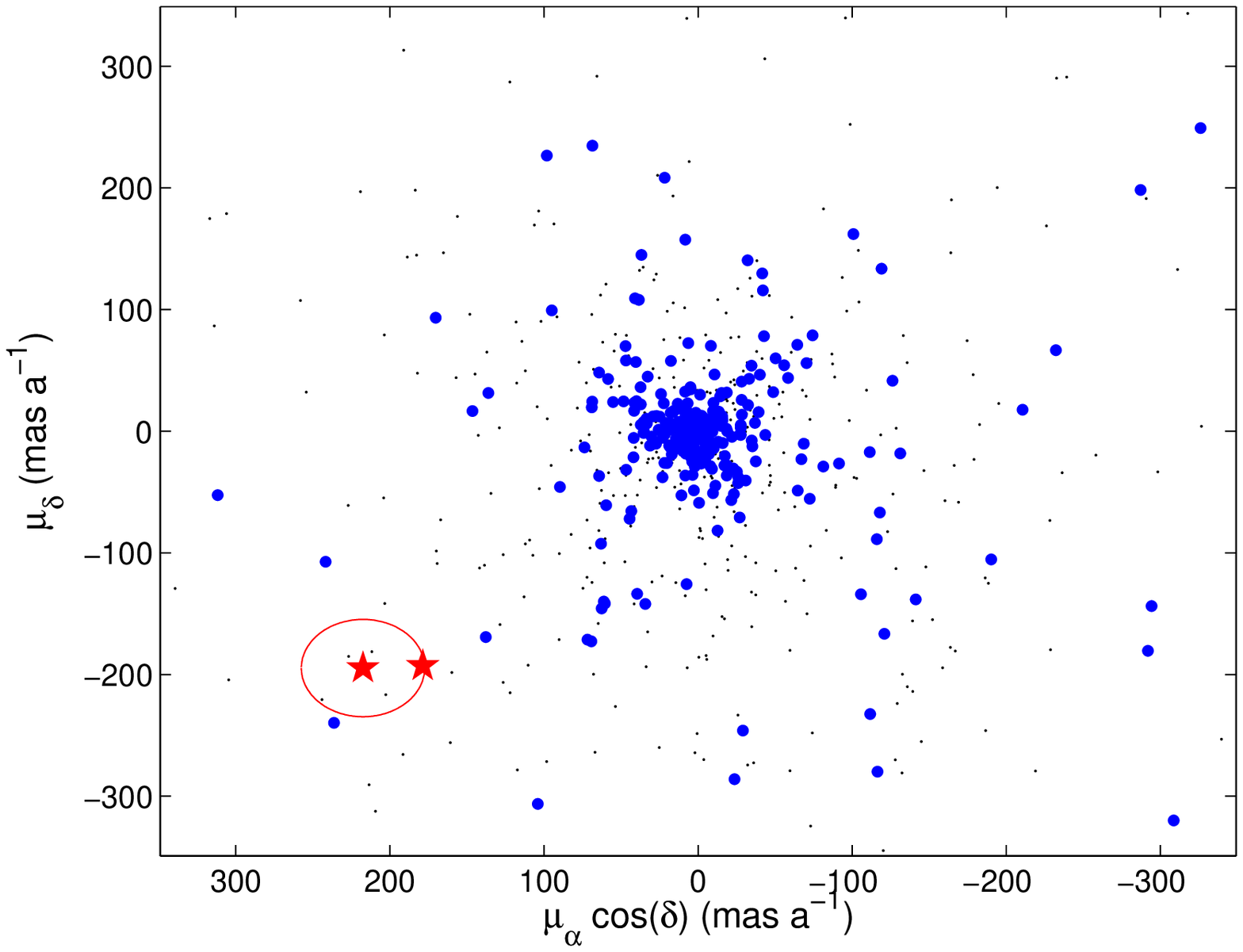}
\plottwo{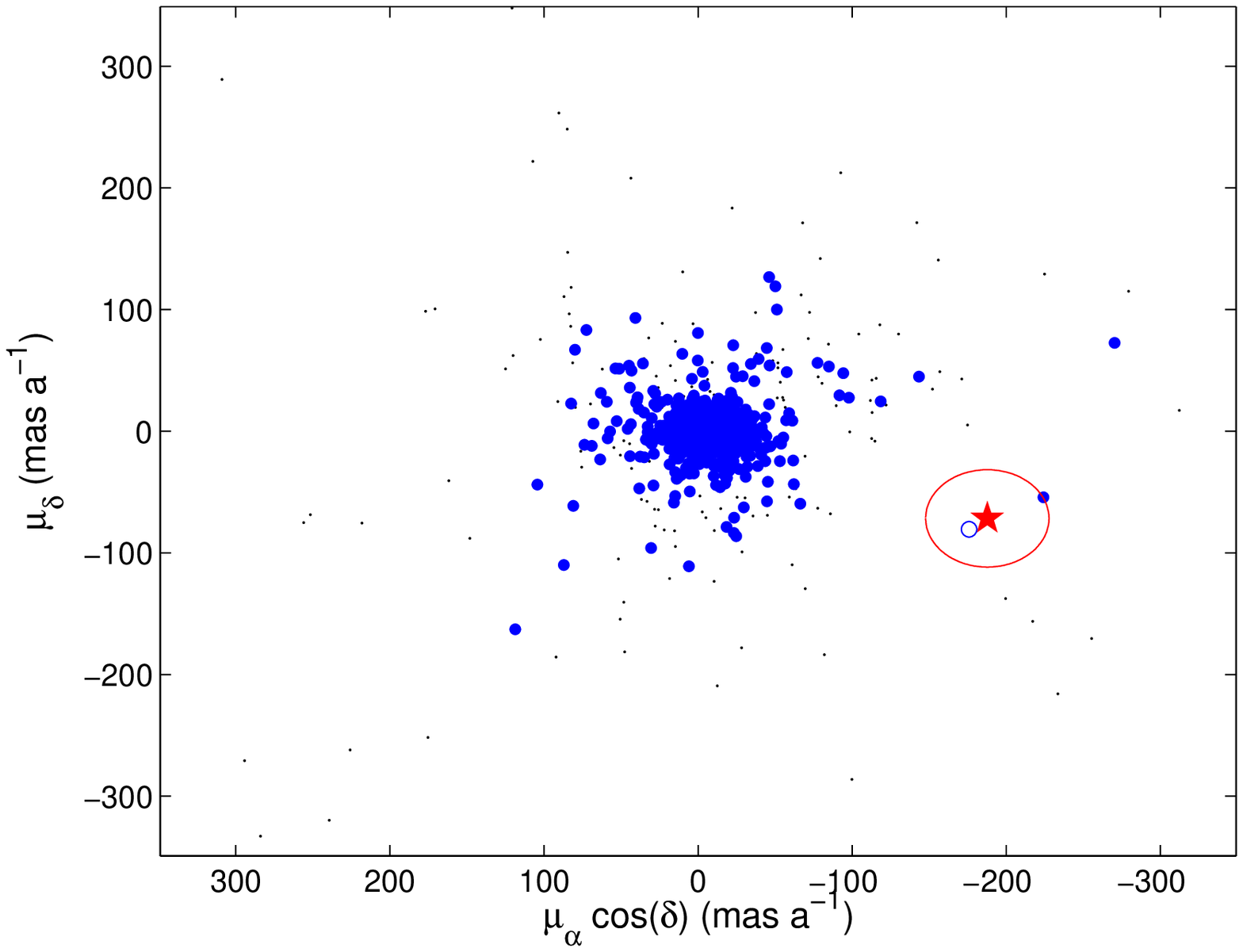}{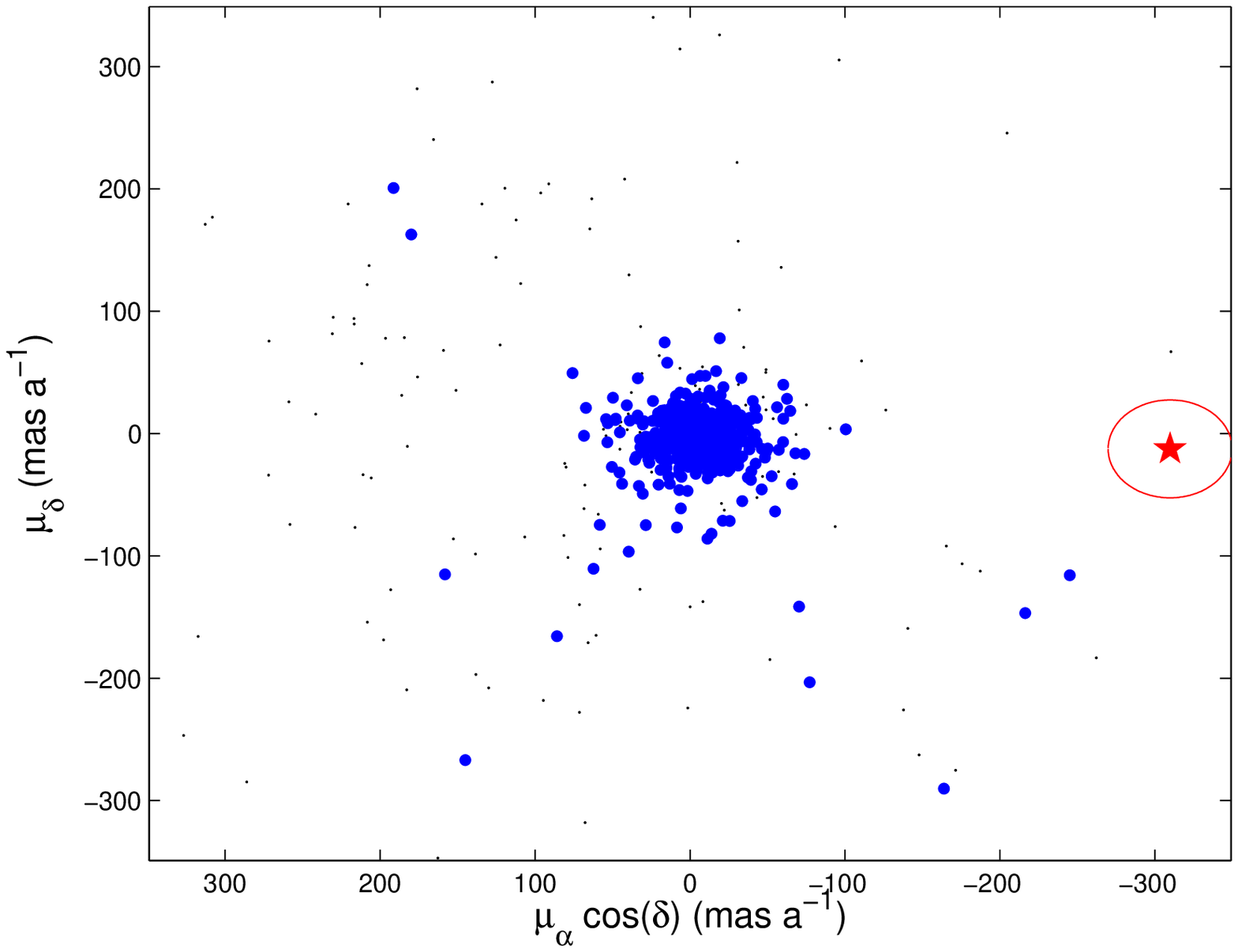}
\caption{Proper-motion ($\mu_\delta$ vs. $\mu_\alpha \cos{\delta}$) diagrams of
four representative VLM dwarfs under study. 
Filled stars: VLM dwarfs and their proper-motion companions.
Small filled circles: Background sources with detections in at least three
SuperCOSMOS passbands. 
Tiny dots: All the detections, including spurious.
 Big open ellipses: 40\,mas\,a$^{-1}$-radius circles centered on the VLM
dwarfs. 
The scales are identical in the four diagrams.
{\em Top left:} K\"o~2AB (LP 655--23 and 2M0430--08; the small filled
circle close to K\"o~2AB is an artefact in the glare of the nearby star
BD--02~912);
{\em top right:} K\"o~3A--BC (HD~221356 and 2M2331--04AB);  
{\em bottom left:} 2M1339--17 and the probable non-common proper motion
companion LP~679--39, marked with a small open circle;
{\em bottom right:} Kelu~1~AB.
The proper-motion diagram of K\"o~1AB is in Caballero~(2007).
\label{muramude}}
\end{figure}

The survey procedure was similar to that in Caballero (2007).
I downloaded the astrometric and photometric SSA data of all the sources in a
10-arcmin radius centered in each field dwarf and searched for stars or
brown dwarfs with similar proper motions to those of the main targets.
The threshold, $\Delta \overline{\mu}$, of the ``similarity'' was at about four
times the error in the proper motion of the programme field dwarfs 
($\Delta \overline{\mu} \approx 4 \delta \mu$, where $\delta \mu =
(\delta^2 \mu_\alpha \cos{\delta} + \delta^2 \mu_\delta)^{1/2}$). 
The error $\delta \mu$ increased for faint objects with late spectral
types and large proper motions. 
Once a common-proper-motion candidate was selected, it was astrometrically
followed up using multi-epoch digitized plates from POSS I Red, UKST Blue, Red,
and Infrared, and data from 2MASS and DENIS catalogues (and the {\em Spitzer}
Science Archive, if available).
Time base-line was typically from first epochs 1950--1954 to last epochs
1999--2000, covering about half a century.
Spurious SSA detections at only two blue optical bands without near-infrared
counterparts were discarded from the study. 
The total survey area was 15.1\,deg$^2$.
Fig.~\ref{muramude} illustrates the proper-motion diagrams of  four
representative VLM dwarfs under study.

\section{Results}

In the Koenigstuhl sample, there are 15 known tight binary and triple
systems unresolved in the SSA images (neither in the 2MASS data). 
They are marked with ``AB (C)'' in Table~\ref{investigated}.
Besides, there are only five previously-known resolved common-proper-motion
multiple systems: $\alpha$~Cen~AB + Proxima Centauri, V1054~Oph ABCDE,
G~124--62~A-- BC ($r \sim$ 1500\,AU; BC: \object{DENIS--P
J144137.3-094559}~AB; Seifahrt, Guenther \& Neuh\"auser 2005), GJ~1001 A--
BC ($r \sim$ 180\,AU; Goldman et al. 1999), and 2M0126--50AB.

\subsection{Koenigstuhl 1, 2, and 3}

Three new common-proper motion systems have arised from the Koenigstuhl survey.
Their basic properties  ($\rho$, $\theta$, $d$, $r$, $M_1$, $M_2$) are
summarized in Table~\ref{new}. 
The uncertainties in the determination of the common proper motions of the
two components in the three systems, measured with the value
$\frac{\sigma_\rho}{\Delta {\rm t}~/~\mu}$ (where $\sigma_\rho$ is the standard
deviation of the mean angular separation, $\Delta {\rm t}$ is the time baseline, and $\mu$ is the modulus of the mean proper motion) are at the level of only
1.1--3.3\,\%. 
 False-color images centered on two of the new common-proper motion systems
are shown in Figs.~\ref{rgb2} and~\ref{rgb3}.

\subsubsection{Koenigstuhl~1~AB (K\"o~1AB)}

K\"o~1AB, formed by \object{LEHPM 494} and \object{DENIS--P~J0021.0--4244}, was
presented in Caballero (2007).  
In this work, I provide a new imaging epoch obtained with the IRAC
instrument onboard the {\em Spitzer} Space Telescope.
I downloaded the images of the four channels, taken on J2003.970 (four years
after the last epoch in Caballero 2007) and performed standard astrometry.
The new measurement of the angular separation, of 77.74$\pm$0.10\,arcsec,
perfectly agrees with what was expected. 
I compute more accurate average separation  and position angle of K\"o~1AB,
given in Table~\ref{new}. 
The expected semimajor axis of the parallax ellipse is $\sim$0.04\,arcsec.

\subsubsection{Koenigstuhl~2~AB (K\"o~2AB)}

The second new common-proper-motion is formed by LP 655--23 and
\object{2MASS J0430516--084901} (K\"o~2AB).
They maintain a constant angular separation of 19.7$\pm$0.2\,arcsec during six
epochs from J1954.005 and J2000.005. 
%Their proper motions, of about 166\,mas\,a$^{-1}$, are the same with an
%uncertainty of 5\,mas\,a$^{-1}$.
The  VLM field dwarf target was the secondary, an M8.0V at 22.9$\pm$1.9\,pc
(Cruz et al. 2003). 
The primary, LP 655--23, was tabulated in the high proper-motion stars
Luyten-Palomar and New Luyten Two Tenths catalogs (Luyten 1979).
Improved astrometry, identical within the errorbars to the one presented here,
was published by Salim \& Gould (2003). 
None of them has been further investigated.
Assuming that the binary is older than 1\,Ga, the NextGen98 models of Baraffe et
al. (1998) and the Dusty00 models of Chabrier et al. (2000) provide masses of
0.26$\pm$0.04 and 0.086$\pm$0.004\,$M_\odot$ for the primary and the secondary,
respectively ($q$ = 0.33$\pm$0.05). 
The colors and the theoretical effective temperature from the models of
LP~655--23 correspond to early-M spectral type. 
Using the distance estimate by Cruz et al. (2003), both M dwarfs are separated
by 450$\pm$40\,AU.
This value makes K\"o~2AB to be the second widest system in the field
with $M_1+M_2<$ 0.4\,$M_\odot$ after K\"o~1AB and 2M0126--50AB, and together with
the M4.5V+L6.0V binary \object{LP~261--75} + \object{2MASS~J09510549+3558021}
($\rho$ = 450$\pm$120\,AU; Reid \& Walkowicz 2006)

\subsubsection{Koenigstuhl~3~A--BC (K\"o~3A--BC)}

The third  and last new common-proper motion system, K\"o~3A-- BC, is
formed by the F8V star \object{HD~221356}~A and the M8.0V+L3.0V binary
\object{HD~221356~BC} (BC: \object{2MASS J23310161--0406193}AB).  
In the discovery paper of 2M2331--04 (as a single object), Gizis et al. (2000)
reported that the derived photometric distance to the M8.0V was consistent with
the Hipparcos distance to the nearby star HD~221356.
However, the proper motion of the secondary tabulated by them,
(+401,~--231) mas\,a$^{-1}$, clearly deviated from the Hipparcos proper motion
of the F8V, (+178.6$\pm$1.0,~--192.8$\pm$0.8) mas\,a$^{-1}$.
The M8.0V was afterwards found to be an 0.573-arcsec double by Gizis et
al.~(2003).  

During the astrometric follow up, I have used six epochs from J1951.583 to
J1999.882, and measured the mean separation between HD~221356 and 2M2331--04AB
at $\rho =$ 451.8$\pm$0.4\,arcsec (the photo-centroid of the primary was
computed using their spikes as a reference). 
The projected physical separation of 11\,900$\pm$300\,AU makes the triple the
widest star system with an L-type component (it is 3.3 times wider than the
\object{$\eta$~CrB} ABC, the formerly widest such system, which is formed by an
L8V secondary and a G1V+G3V spectroscopic-binary primary;  Kirkpatrick et
al.~2001).

The measurement of the common proper motion, with an uncertainty of only
9\,mas\,a$^{-1}$, is very important because it helps to constraint the
properties of the VLM binary HD~221356~ BC/\-2M2331--04AB/\-
K\"o~BC.   
Using the Hipparcos trigonometric parallax of HD~221356~A, the age of 
5.7$^{+9.0}_{-0.2}$\,Ga tabulated by Nordstr\"om et al. (2004), the combined
2MASS $J$ magnitude of K\"o~3BC (Cutri et al. 2003), their $\rho$ and
$\Delta J$ given by Gizis et al. (2003), and the Dusty00 models I have
determined new accurate theoretical masses for the M8.0V+L3.0V binary
(given in Table~\ref{new}). 
The L3.0V has an estimated mass larger than previously estimated.
 The errorbars in the masses only account for the uncertainties in the
distance, age and $J$-band magnitudes, but not for the systematic errors of the 
theoretical models. 
The determination of the dynamical masses of K\"o~3BC through astrometric and
radial-velocity monitoring will help to estimate those systematic errors.  
The orbital period of K\"o~3BC, $P \sim$ 146\,a ($a \sim$
15.0\,AU; I assume a circular, face on, orbit, and adoption of the separation as
the semimajor axis of the orbit), is quite similar to that predicted by Gizis
et al. (2003).  
The orbital period of the binary surrounding the F8V is a bit larger than 1\,Ma.
Finally, the metallicity of the primary and, therefore, of the system is
also known ([M/H] = --0.23; Karata\c{s}, Billir \& Schuster 2005), which may
help to further spectral classification of the L3.0V component
(Kirkpatrick~2005).

\subsection{Probable background non-companions}
\label{back}

I have found 14 stars at angular separations less than 10\,arcmin to the
investigated dwarfs (12\,000\,AU at a typical heliocentric distance of 20\,pc)
with proper motions within the 4$\Delta \overline{\mu}$ threshold and that seem
to be  background stars with spectral types earlier than M5V.  
Their basic data are given in Table~\ref{blue}.
BD--20 3682 is an early-F star located at 200$\pm$70\,pc to the Sun  from
{\em Hipparcos} parallax and at 7.0\,arcmin to 2M1237--21, which in contrast is
an M6 dwarf at only 32$\pm$6\,pc (Cruz et al. 2003). 
BD--20 3682 was classified as a low-metallicity subdwarf by Ryan \& Norris
(1991).
HD~117332, the brightest background non-companion, is a G0-type star whose
X-ray counterpart was detected by Schwope et al. (2000).
It is located far beyond the 27.0$\pm$2.2\,pc estimated by Cruz et al. (2003)
for 2M1330--04. 
Of the remaining 12 stars, six are fainter than the VLM target dwarfs, but have
bluer optical-near infrared colors (e.g. $I-K_{\rm s} \lesssim$ 1.6\,mag), on
the contrary to what was expected if they formed a common-proper motion pair.
Other three stars are brighter than the VLM targets, but the distances roughly
estimated from their colors and magnitudes do not match those of the VLM
dwarfs.

I have made astrometric follow up of the three remaining companion
candidates, that are brighter than their respective VLT dwarfs:
2MASS~J012704.7--501711, LP~679--39 and LP~798--19. 
On the one hand, LP~679--39 is a background G:-type star (SIMBAD)
whose projected physical separation of 6.4\,arcmin to 2M1413--04 varied
6\,arcsec in a time base-line of 42\,a and, therefore, they do not share a
common proper motion. 
On the other hand, I failed to ascertain the common-proper motion status of
LP~798--19 and 2M1339--17 ($\rho $ = 9.489$\pm$0.010\,arcmin, $\theta$ =
342.88$\pm$0.09\,deg) and of 2MASS~J012704.7--5017112 and 2M0126--50AB ($\rho$ =
5.656$\pm$0.006\,arcmin, $\theta$ = 344.93$\pm$0.06\,deg).    
2M1339--17 is an M7.5V located at 31$\pm$3\,pc (Cruz et al. 2003), while
LP~798--19 seems to be an early M at 30--40\,pc, based on their optical and 
near-infrared magnitudes. 
 2M0126--50AB is the wide equal-mass binary found by Artigau et al. (2007),
with a photometric distance of $d \sim$ 62\,pc, while 2MASS~J012704.7--5017112,
with a red color $I-K_{\rm s} \sim$ 2.2\,mag and about 2\,mag brighter in $J$
than 2M0126--50AB, is investigated here for the first time.
I have measured marginal  variations of $\Delta \rho \sim$ 1\,arcsec of the
two systems during 43- and 20-year base-lines.
Additional imaging epochs are needed to discard or confirm their common
proper motions.

\subsection{Miscellanea}

As a by-product of the survey, I have measured for the first time the proper
motions of 76 VLM field dwarfs (marked with ``(1)'' in
Table~\ref{investigated}).
 Accurate, homogeneous coordinates are also provided for the 173 dwarfs and
8 resolved proper-motion companions.

I have determined the mean angular separation between 2M0126--50A and B
(Artigau et al. 2007) at 81.93$\pm$0.18\,arcsec, constant within the
uncertainties during my time baseline of~18.0\,a (2M0126--50B is not
visible in the UKST $B_J$~digitization).   

Also, the double 2M0429--31AB (M7.5V+L1.0V -- Cruz et al. 2003; Siegler et
al. 2005) is at only 7.2\,arcsec to the faint X-ray source
\object{1RXS~J042918.9--312401} (Voges et al. 2000), suggesting~relationship.

\section{The frequency of wide very low-mass binaries}

Of the 173 investigated VLM dwarfs, 13 have large $\delta
\mu$-to-$\mu$ ratios (marked with ``(3)'' in Table~\ref{investigated}), which
prevented from searching common-proper motion companios surrounding them.
Therefore, 160 dwarfs remain for statistical purposes.
Taking into account the 15 unresolved systems, the five previously-known
resolved systems, and the three new Koenigstuhl systems, then the 
frequency of multiplicity in the magnitude-limited sample of VLM dwarfs in
the spectral-type interval M5.5V--L8.0V is $\gtrsim$ 14\,\%.
This value is a lower limit because most of the programme targets have not been
yet investigated with high-spatial-resolution facilities. 
I refer to Close et al. (2003), Siegler et al. (2005), Burgasser et al.
(2005), and references therein to find accurate frequencies of close multiples
($r <$ 30\,AU) at the 10--30\,\% level. 
 These values must be compared to the upper limit of relatively wide
companion frequency at 2--31\,arcsec to M7--L8 dwarfs recently determined by
Allen et al. (2007), of 2.3\,\%.
In contrast to these works, the Koenigstuhl survey is the only one able to study
the frequency of very wide multiples ($r >$ 100\,AU)  up to
6000--30\,000\,AU (at heliocentric distances $d$ = 5--50\,pc).
In the aforementioned spectral-type interval, the minimum frequency of VLM
dwarfs in wide multiple systems is as low as 5.0$\pm$1.8\,\% (8 of 160;
Poissonian errors).  
The actual frequency could be larger because this survey is not sensitive to the
detection of very faint~companions.

There are only two wide binaries in my survey with mass ratio $q >$ 0.5,
K\"o~1AB and 2M0126--50AB (the other known field wide equal-mass VLM
binary, DE0551--44AB, although it is in the Southern Hemisphere, is too faint
for the magnitude-limited Koenigstuhl survey).
The frequency of wide equal-mass VLM binaries is, therefore,
1.2$\pm$0.9\,\%.   
 Despite the fact that it is not clear whether the origin of the wide
separations between K\"o~1A and B and 2M0126--50A and B resides on the formation
mechanism or in the gravitational tidal disruption within the Galactic disk (or
in both of them), my survey has confirmed the low frequency of wide equal-mass
VLM~binaries.
Further theoretical studies of formation in collapsing molecular clouds and of
the interaction of low binding-energy binaries with the gravitational field of
the Galactic disk must account this low frequency.

To derive a more accurate frequency of wide equal-mass VLM binaries, the
Koenigstuhl survey should be complemented in the future with new very wide
photometric and astrometric searches in both Southern and Northern Hemispheres.

\section{Summary}

 I have investigated 173 very low-mass stars and brown dwarfs during a
proper-motion survey of resolved binary and multiple systems with very low-mass
components, named Koenigstuhl. 
The studied field dwarfs have spectral types $>$ M5.0V, magnitudes $J
\lesssim$ 14.5\,mag, and declinations $\delta <$ +3\,deg. 
I looked for common-proper companions within a radius of 10\,arcmin centered on
the dwarfs using astrometric data from the SuperCOSMOS Science Archive.
Of the investigated very low-mass dwarfs, 160 could actually be searched.
I firstly provide the proper motions of 76 dwarfs.

I have identified five previously known wide multiples, confirmed the
common-proper motion of two wide very low-mass binaries with mass ratio $q >$
0.5 (Koenigstuhl 1~AB and 2M0126--50AB), and measured for the first time the
common-proper motion of two new wide systems containing very low-mass
components, Koenigstuhl~2~AB and 3~A--BC. 
Koenigstuhl 2~AB is formed by the early-M, high proper-motion star LP~655--23
and the M8.0V dwarf 2M0430--08.
Their low total mass ($M_1 + M_2 \approx$ 0.35\,$M_\odot$) and relatively large
separation ($\rho$ = 450$\pm$40\,AU) and mass ratio ($q$ = 0.33$\pm$0.05) make
the system to be one of the lowest-mass, widest binaries yet found.
The components of Koenigstuhl~3~A--BC are the F8V star HD~221356 and the
M8.0V+L3.0V tight binary 2M2331--04AB.
They are separated by $\sim$7.5\,arcmin ($\sim$12\,000\,AU at the {\em
Hipparcos} distance of the primary), which makes Koenigstuhl~3~A--BC to be by
far the widest system containing an L-type dwarf.
The knowledge of the basic properties of the primary (distance, age,
metallicity) and, therefore, of the very low-mass binary companion, will allow
to test theoretical models and classification schemes of ultracool dwarfs with
very late spectral~types. 

Finally, I have determined the minimum frequency of field wide multiples ($r >$
100\,AU) with very low-mass components at 5.0$\pm$1.8\,\% and the frequency of
field wide very low-mass components binaries with mass ratios $q >$ 0.5 at
1.2$\pm$0.9\,\%.

\acknowledgments

I thank T. J. Henry for his valuable refereeing.
I also thank M. R. Bate, {J. E. Gizis}, B. Goldman, R. Mundt, and N. Phan-Bao
for helpful comments. 
I have used IRAF, 
the M, L, and T dwarf  compendium\footnote{\tt http://DwarfArchives.org}, 
the RECONS (Research Consortium on Nearby Stars) List of the Nearest 100 Stellar
Systems\footnote{\tt http://www.chara.gsu.edu/RECONS/}, 
the Very Low Mass Binaries Archive\footnote{\tt
http://paperclip.as.arizona.edu/$\sim$nsiegler/VLM\_binaries},
the Extrasolar Planets Encyclopaedia\footnote{\tt http://exoplanet.eu/}, 
the Two-Micron All Sky Survey, the Deep Near Infrared Survey of the Southern
Sky, the USNO-B1 and NOMAD catalogues, the SuperCOSMOS and {\em Spitzer} Science
Archives, and the SIMBAD database. 
Partial financial support was provided by the Spanish Ministerio de
Ciencia y Tecnolog\'{\i}a project AYA2004--00253 of the Plan Nacional de
Astronom\'{\i}a y Astrof\'{\i}sica.

\clearpage

\begin{deluxetable}{lllcccccc}
\tabletypesize{\scriptsize}
\tablecaption{New common-proper motion systems identified in the Koenigstuhl
survey.\label{new}}  
\tablewidth{0pt}
\tablehead{
\colhead{Name} & 
\colhead{Primary} & 
\colhead{Secondary} & 
\colhead{$\rho$} & 
\colhead{$\theta$} & 
\colhead{$d^a$} & 
\colhead{$r$} & 
\colhead{$M_1^b$} & 
\colhead{$M_2^b$} 
\\
 & 
 & 
 & 
\colhead{(arcmin)} & 
\colhead{(deg)} &
\colhead{(pc)} &
\colhead{(AU)} &
\colhead{($M_\odot$)} &
\colhead{($M_\odot$)} 
}
\startdata
K\"o 1AB	& LEHPM 494	& DE0021--42	& 1.2956$\pm$0.0012	& 316.97$\pm$0.08	& 23$\pm$2	& 1800$\pm$170  & 0.103$\pm$0.006	& 0.079$\pm$0.004 \\
K\"o 2AB	& LP 655--23	& 2M0430--08	&  0.328$\pm$0.004      & 339.9$\pm$0.4		& 22.9$\pm$1.9  &  450$\pm$40   & 0.26$\pm$0.04		& 0.086$\pm$0.004 \\
K\"o 3A--BC	& HD 221356	& 2M2331--04AB	&  7.530$\pm$0.007      & 261.77$\pm$0.06	& 26.2$\pm$0.6  &11900$\pm$300  & 1.02$^{+0.07}_{-0.06}$& 0.088$\pm$0.002 (B) \\
		& 		& 		&  		      	&			&   		&  		& 			& 0.072$\pm$0.001 (C) \\
\enddata
\tablenotetext{a}{Errors in distance estimates have been adopted from the literature.}  
\tablenotetext{b}{Mass errors are from theoretical fits to available data, and are not realistic.}  
\end{deluxetable}

\clearpage

\begin{deluxetable}{lcccccccc}
\tabletypesize{\scriptsize}
\tablecaption{Probable non-common proper motion companions to the investigated
dwarfs.\label{blue}} 
\tablewidth{0pt}
\tablehead{
\colhead{Name} & 
\colhead{VLM} & 
\colhead{$\mu_\alpha \cos{\delta}$} & 
\colhead{$\mu_\delta$} & 
\colhead{$B$} &
\colhead{$R$} &
\colhead{$I$} &
\colhead{$J$} &
\colhead{$K_{\rm s}$} \\
 & 
\colhead{dwarf} & 
\colhead{(mas\,a$^{-1}$)} & 
\colhead{(mas\,a$^{-1}$)} & 
\colhead{(mag)} &
\colhead{(mag)} &
\colhead{(mag)} &
\colhead{(mag)} &
\colhead{(mag)} 
}
\startdata
\object{G 271--43}$^{a}$      		& DE0120--07    &  +20$\pm$30   &--180$\pm$30  	& $\sim$15.7 	& $\sim$14.0 	& 13.74$\pm$0.03& 12.87$\pm$0.02& 12.19$\pm$0.03\\ %    5.3 Brighter but much bluer
\object{2MASS J012704.7--501711}  	& 2M0126--50AB  & +135$\pm$12	& --12$\pm$9	& $\sim$17.8 	& $\sim$15.6 	& $\sim$14.0 	& 12.57$\pm$0.02& 11.79$\pm$0.02\\ %    5.6 $^{b}$		      
\object{2MASS J033411.0--212412}      	& 2M0334--21    & +139$\pm$12	& --12$\pm$11	& $\sim$21.1 	& $\sim$18.7 	& $\sim$16.6 	& 15.25$\pm$0.04& 14.41$\pm$0.09\\ % X  6.3 Fainter and bluer		      
\object{2MASS J095210.2--193029}      	& DE0952--19    & --71$\pm$7	& --96$\pm$7	& $\sim$18.7 	& $\sim$16.4 	& $\sim$15.0 	& 14.11$\pm$0.04& 13.39$\pm$0.04\\ % Y  6.5 Fainter and much bluer	      
\object{BD--20 3682}$^{a,b}$      	& 2M1237--21  &--177.1$\pm$1.8 &--43.0$\pm$1.3 	& 11.09$\pm$0.07& $\sim$9.9 	& 10.14$\pm$0.02&  9.72$\pm$0.03&  9.39$\pm$0.02\\ % Za 7.0 Brighter but much bluer, VI:       
\object{2MASS J123723.7--210939}      	& 2M1237--21    &--201$\pm$9	& --47$\pm$7	& $\sim$20.2 	& $\sim$18.0 	& 16.21$\pm$0.06& 14.74$\pm$0.04& 13.90$\pm$0.05\\ % Zb 8.1 Fainter and bluer		       
\object{2MASS J123758.2--211521}      	& 2M1237--21    &--259$\pm$12	& --37$\pm$10	& $\sim$20.1 	& $\sim$19.5 	& 17.06$\pm$0.13& 15.77$\pm$0.05& 14.90$\pm$0.11\\ % Zc 7.6 Fainter and bluer		       
\object{HD 117332}$^{a,b}$      	& 2M1330--04    & --34$\pm$3  	&   +4$\pm$2 	& 10.28$\pm$0.04& $\sim$9.0 	& $\sim$8.9	&  8.10$\pm$0.04&  7.65$\pm$0.02\\ %    4.8 Brighter but much bluer
\object{2MASS J132947.9--050125}      	& 2M1330--04    & --58$\pm$7    &  --3$\pm$6  	& $\sim$17.1 	& $\sim$15.2 	& $\sim$14.4	& 12.69$\pm$0.03& 12.52$\pm$0.03\\ %    8.8 Brighter and bluer
\object{LP 798--19}$^{a}$   		& 2M1339--17    &--224$\pm$9    & --54$\pm$11	& $\sim$14.8	& $\sim$12.8	& 11.34$\pm$0.03& 10.00$\pm$0.02&  9.21$\pm$0.02\\ %    9.4 Salim & Gould (2003) $^{b}$ 13 39 38.2 & --18 04 09
\object{2MASS J135751.0--143458}      	& 2M1357--14    & --58$\pm$7    &--118$\pm$6  	& $\sim$18.8 	& $\sim$17.4 	& $\sim$16.8	& 15.99$\pm$0.08& 15.34$\pm$0.19\\ %    9.5 Much bluer
\object{LP 679--39}$^{a}$      		& 2M1413--04    &--149$\pm$9    &--132$\pm$7  	& $\sim$13.4 	& $\sim$11.6 	& 12.01$\pm$0.02& 11.13$\pm$0.03& 10.37$\pm$0.02\\ %    6.4 Brighter and bluer $^{b}$
\object{2MASS J220659.4--204323}      	& DE2206--20AB  &  +34$\pm$14	& --30$\pm$20	& $\sim$13.7 	& $\sim$11.8	& 11.96$\pm$0.02& 11.18$\pm$0.02& 10.55$\pm$0.02\\ % U  9.3 Brighter but much bluer		 
\object{2MASS J230702.5--050234}      	& 2M2306--05    & --55$\pm$8	& --93$\pm$8	& $\sim$20.7 	& $\sim$17.8	& $\sim$16.8 	& 15.41$\pm$0.06& 14.78$\pm$0.10\\ % V  8.2 Fainter and bluer		      
\enddata
\tablenotetext{a}{J2000 coordinates. 
G 271--43: 01~21~10.0 --07~39~21;  
BD--20 3682: 12~36~59.3 --21~20~38; 
HD 117332: 13~29~43.2 --04~54~22; 
LP 798--19: 13~39~38.2 --18~04~09; 
LP 679--39: 14~14~21.8 --04~54~16.}  
\tablenotetext{b}{$B$ magnitudes and proper motions from the Hipparcos
catalogue.}  
%\tablenotetext{b}{Discussed in Section~\ref{back}.}  
\end{deluxetable}

\clearpage

\begin{deluxetable}{llccccccc}
\setlength{\tabcolsep}{0.03in}
\tabletypesize{\scriptsize}
\tablecaption{Investigated very low-mass dwarfs and proper-motion
companions.\label{investigated}}  
\tablewidth{0pt}
\tablehead{
\colhead{Name} & 
 & 
\colhead{Alternative} &
\colhead{$\alpha$} &
\colhead{$\delta$} &
\colhead{$\mu_\alpha \cos{\delta}$} & 
\colhead{$\mu_\delta$} & 
\colhead{Sp.} &
\colhead{Remarks} 
\\
 & 
 & 
\colhead{name} &
\colhead{(J2000)} &
\colhead{(J2000)} &
\colhead{(mas\,a$^{-1}$)} & 
\colhead{(mas\,a$^{-1}$)} & 
\colhead{type} &
\colhead{($^a$)}
}
\startdata
\object{LP 584--4} 		&  	&			& 00 02 06.2	&  +01 15 36	& +480$\pm$20	&   +30$\pm$20	& M6.5V	    & 			\\     
\object{GJ 1001} 		& A 	&			& 00 04 36.4	& --40 44 02	& +770$\pm$60	&--1600$\pm$40	& M3.5V	    & $(2)$		\\     
				& BC    &			& 00 04 34.8    & --40 44 06    & +710$\pm$90	&--1580$\pm$80  & L4.5V+... & $(2)$	        \\ 
\object{GJ 1002}		&       &			& 00 06 43.3    & --07 32 15    &--800$\pm$60	&--1920$\pm$70  & M5.5V     & $(2)$	        \\ 
\object{LP 825--35} 		&  	& LEHPM 485		& 00 20 23.2	& --23 46 05	& +321$\pm$11	&  --73$\pm$10	& M6.0V	    & 			\\     
Koenigstuhl 1 			& A     & LEHPM 494		& 00 21 10.4	& --42 45 40	& +268$\pm$10	&  --21$\pm$8	& M6.0:V    &		        \\ 
				& B     & DE0021--42		& 00 21 05.7	& --42 44 43	& +270$\pm$11	&    +4$\pm$10  & M9.5V     &		        \\ 
\object{DY Psc}			&       & BRI B0021--0214	& 00 24 24.6	& --01 58 20	& --80$\pm$7	&  +137$\pm$7	& M9.5V     &		        \\ 
\object{GJ 2005}		& ABC   & LP 881--64		& 00 24 44.2	& --27 08 24	& --80$\pm$50	&  +610$\pm$70  & M5.5V+... &		        \\ 
\object{DENIS--P J004135.3--562112}&    &			& 00 41 35.4	& --56 21 13	& +108$\pm$10	&  --63$\pm$8   & M7.5V     & 			\\ 
\object{2MASS J00492677--0635467}&      &			& 00 49 27.9	& --06 35 40	&--111$\pm$8	& --460$\pm$30  & M8.5V     & 			\\ % $(1')$
\object{RG 0050--2722}		&       &			& 00 52 54.7	& --27 06 00	& +229$\pm$17	& --332$\pm$16  & M8.0V     & 	        	\\ 
\object{LP 938--71}		&       & LHS 132		& 01 02 51.0	& --37 37 44	&+1480$\pm$30	&  +200$\pm$30  & M8.0:V    &		        \\ 
\object{DENIS J010311.9--535143}&  	&			& 01 03 12.0	& --53 51 43	& --89$\pm$7	& --204$\pm$5	& M5.5V	    & 			\\     
\object{SSSPM J0109-5101}	&       & 			& 01 09 01.5	& --51 00 50	& +207$\pm$12	&   +87$\pm$11  & M8.5V     &		        \\ 
\object{LP 647--13}		&       & NLTT 3868		& 01 09 51.2	& --03 43 26	& +380$\pm$50	&   +20$\pm$50  & M9.0V     &		        \\ 
\object{DENIS--P J012049.1--074103}&    & 			& 01 20 49.1	& --07 41 03	&   +4$\pm$9	& --153$\pm$9	& M8.0V     & 			\\ % $(1')$
\object{LEHPM 1505}		&    	& SSSPM J0124--4240	& 01 24	59.1	& --42 40 07	&--141$\pm$9	& --227$\pm$8	& M7.0V     & 			\\ 
\object{2MASS J01265549--5022388}& A   	& 2M0126--50A		& 01 26 55.5	& --50 22 39	& +136$\pm$15	&  --47$\pm$14	& M6.5V     & 			\\ 
				& B   	& 2M0126--50B		& 01 27 02.8	& --50 23 21	& +180$\pm$170	&  +160$\pm$160	& M8.0V     & $(3)$		\\ 
\object{BL Cet} + \object{UV Cet}&      & GJ 65 AB		& 01 39 01.5	& --17 57 02	&+3295$\pm$5	&  +563$\pm$5	& M5.5Ve+...& $(4)$	        \\ 
\object{LEHPM 1781}		&       & 			& 01 41 14.8	& --24 17 31	&--145$\pm$13	& --307$\pm$12  & M7.5V     &		        \\ 
\object{DENIS J014431.8--460432}&  	&			& 01 44 31.9	& --46 04 32	& +112$\pm$9	&  --47$\pm$8	& M5.5V	    & 			\\    
\object{DENIS--P J014543.4--372959}&  	&			& 01 45 43.5	& --37 29 59	& +175$\pm$12	&  --66$\pm$11	& M7.5V	    &  			\\    
\object{2MASS J01483864-3024396}&       &			& 01 48 38.6	& --30 24 40	& --88$\pm$11	&   +44$\pm$10  & M7.5V     & $(1)$	        \\ 
\object{LP 649--72}		&       & LHS 1363		& 02 14 12.5	& --03 57 43	& +490$\pm$50	& --130$\pm$60  & M6.5V     &		        \\ 
\object{LP 649--93}		&       & PB 9141		& 02 18 57.9	& --06 17 50	& +374$\pm$19	&  --91$\pm$18  & M8.0V     & $(2)$	        \\ 
\object{2MASS J02192807--1938416}&      &			& 02 19 28.0	& --19 38 41	& +221$\pm$18	& --132$\pm$17  & L0.0V     & $(1)$	        \\ 
\object{LP 771--21}		&       & BR B0246--1703	& 02 48 41.0	& --16 51 22	&  +22$\pm$15	& --299$\pm$14  & M8.0V     &		        \\ 
\object{LP 651--17}		&       & LHS 1450		& 02 50 02.4	& --08 08 42	& +590$\pm$20	&  +110$\pm$20  & M5.5V     &		        \\ 
\object{2MASS J02511490--0352459}&      &			& 02 51 14.9	& --03 52 46	&+1000$\pm$200  &--1800$\pm$200 & L3.0V     & 			\\ % $(1')$
\object{DENIS--P J025344.4--795913}&    &			& 02 53	44.5	& --79 59 13	&  +71$\pm$9  	&  +103$\pm$9 	& M5.5V     & 			\\ 
\object{DENIS--P J0255.0--4700} &    	&			& 02 55 03.6	& --47 00 51	&+1060$\pm$50	& --630$\pm$50  & L8.0V     & 			\\ % $(1')$
\object{LEHPM 3070}		&       &			& 03 06 11.6	& --36 47 53	&   +0$\pm$180  & --570$\pm$170 & M8.5V     &		        \\ 
\object{DENIS--P J031225.1+002158}&     &			& 03 12	25.1	&  +00 21 58	& +178$\pm$18	&  --40$\pm$17	& M5.5V     & 	        	\\ 
\object{2MASS J03144011--0450316}&      &			& 03 14 40.1	& --04 50 32	& --86$\pm$7	& --101$\pm$7	& M7.5V     & $(1)$	        \\ 
\object{2MASS J03202839--0446358}&      &			& 03 20 28.4	& --04 46 36	&--190$\pm$60	& --560$\pm$60  & M8.0:V    &		        \\ 
\object{DENIS J032058.8--552015}&  	&			& 03 20 58.9	& --55 20 16	& +297$\pm$8	&  +264$\pm$8	& M5.5V	    & 			\\    
\object{DENIS--P J032426.8--772705}&  	&			& 03 24	26.9	& --77 27 05	& +265$\pm$19	&  +190$\pm$19	& M6.0V	    & 			\\    
\object{LP 888--18}		&       & NLTT 11163		& 03 31 30.2	& --30 42 38	&  +41$\pm$9	& --402$\pm$9	& M7.5V     &		        \\ 
\object{2MASS J03341065--2130343}&      &			& 03 34 10.7	& --21 30 34	& +140$\pm$7	&   --4$\pm$7	& M6.0V     & $(1)$	        \\ 
\object{GJ 1061}		&       & LP 995--46		& 03 35 59.7	& --44 30 45	& +730$\pm$60	& --330$\pm$20  & M5.5V     &		        \\ 
\object{LP 944--20}		&       &			& 03 39 35.2	& --35 25 44	& +290$\pm$12	&  +280$\pm$12  & M9.0V     & $(2)$	        \\ 
\object{LP 593--68}		&       & GJ 3252		& 03 51 00.0	& --00 52 45	&   +1$\pm$12	& --474$\pm$12  & M7.5V     &		        \\ 
\object{2MASS J03521086+0210479}&       &			& 03 52	10.9	&  +02 10 48	& +260$\pm$30	&  +370$\pm$30	& L1.0V     & 	        	\\  % $(1')$
\object{2MASS J03542008--1437388}&      &			& 03 54 20.1	& --14 37 39	&--125$\pm$5	&   +58$\pm$5	& M6.5V     & $(1)$	        \\ 
\object{2MASS J03550477--1032415}&      &			& 03 55 04.8	& --10 32 42	&  +71$\pm$7	&  --35$\pm$7	& M8.5V     & $(1)$	        \\ 
\object{LP 714--37}		& ABC   & 			& 04 10 48.1	& --12 51 42	&--168$\pm$15	& --395$\pm$22  & M6.0V+... & $(2)$	        \\ 
\object{LP 890--2}		&       & NLTT 12812		& 04 13 39.8	& --27 04 29	& +270$\pm$7	&  --31$\pm$7	& M6.0V     &		        \\ 
\object{2MASS J04173745--0800007}&      &			& 04 17 37.5	& --08 00 01	& +670$\pm$70	&  --90$\pm$70  & M7.5V     & $(1)$	        \\ 
\object{2MASSI J0422205--360608}&       &			& 04 22 20.6	& --36 06 08	& +207$\pm$8	&  --40$\pm$8	& M6.5V     & $(1)$	        \\ 
\object{2MASS J04235322--0006587}&	& 			& 04 23 53.2	& --00 06 59	&--130$\pm$160  & --240$\pm$150 & M8.5V     & $(1)$, $(3)$      \\ 
\object{2MASS J04285096--2253227}&      &			& 04 28 51.0	& --22 53 23	&  +97$\pm$13	&  +156$\pm$13  & L0.5V     & 			\\ % $(1')$
\object{2MASS J04291842--3123568}& AB   & 			& 04 29 18.4	& --31 23 57	&  +97$\pm$5	&   +71$\pm$6	& M7.5V+... & $(1)$      	\\ 
\object{LP 655--23}		& A     & NLTT 13422		& 04 30 52.0	& --08 49 19	&   +7$\pm$15	& --161$\pm$12  & M:V	    &		        \\ 
				& B     & 2M0430--08		& 04 30 51.6	& --08 49 01	&  --4$\pm$11	& --160$\pm$11  & M8.0V     & $(1)$	        \\ 
\object{LP 775--31}		&       & NLTT 13580		& 04 35 16.1	& --16 06 58	& +162$\pm$18	&  +313$\pm$20  & M7.5V     &		        \\ 
\object{2MASS J04351455--1414468}&      &			& 04 35 14.6	& --14 14 47	&   +0$\pm$10	&   +11$\pm$10  & young	    & $(1)$, $(3)$      \\ 
\object{DENIS J043627.8--411446}&       &			& 04 36 27.9	& --41 14 46	&  +71$\pm$10	&   +20$\pm$10  & M7.5V     & 			\\ % $(1')$
\object{2MASS J04393407--3235516}&      &			& 04 39 34.1	& --32 35 52	& --97$\pm$5	&   --1$\pm$6	& M6.5V     & $(1)$	        \\ 
\object{LP 655--48}		&       & 			& 04 40 23.2	& --05 30 08	& +335$\pm$7:	&  +115$\pm$8:  & M7.0V     &		        \\ 
\object{2MASS J04451119--0602526}&      &			& 04 45 11.2	& --06 02 53	&  +49$\pm$6	&  --21$\pm$6	& M7.0V     & $(1)$	        \\ 
\object{2MASS J04453237--3642258}&      &			& 04 45 32.4	& --36 42 26	& +520$\pm$70	&   +10$\pm$70  & M9.0:V    & $(1)$	        \\ 
\object{2MASS J04455387--3048204}&      &			& 04 45 53.9	& --30 48 20	& +167$\pm$12	& --424$\pm$12  & L2.0V     & $(1)$	        \\ 
\object{2MASS J04510093--3402150}&      &			& 04 51 00.9	& --34 02 15	&  +94$\pm$17	&  +114$\pm$16  & L0.5V     & $(1)$	        \\ 
\object{2MASS J05023867--3227500}&      &			& 05 02 38.7	& --32 27 50	&  +53$\pm$7	& --175$\pm$7	& M6.5V     & $(1)$	        \\ 
\object{2MASS J05084947--1647167}&      &			& 05 08 49.5	& --16 47 17	&--220$\pm$20	& --360$\pm$20  & M8.0V     & $(1)$	        \\ 
\object{DENIS--P J051737.7--334903}&    &			& 05 17 37.7	& --33 49 03	& +460$\pm$12	& --319$\pm$12  & M8.0V     & $(2)$ 		\\ % $(1')$
\object{2MASS J05233822--1403022}&      &			& 05 23 38.2	& --14 03 02	& +105$\pm$7	&  +158$\pm$7	& L2.5V     & $(1)$	        \\ 
\object{2MASS J05284435--3252228}&      &			& 05 28 44.4	& --32 52 23	& --10$\pm$20	&   +50$\pm$20  & M8.5V     & $(1)$, $(3)$      \\ 
\object{2MASS J06003375--3314268}&      &			& 06 00 33.8	& --33 14 27	& --15$\pm$10	&  +119$\pm$11  & M7.5V     & $(1)$	        \\ 
\object{2MASS J06080232--2944590}&      &			& 06 08 02.3	& --29 44 59	&  +30$\pm$20	&  +100$\pm$20  & M8.5V     & $(1)$	        \\ 
\object{2MASS J06085283--2753583}&      &			& 06 08 52.8	& --27 53 58	&  +30$\pm$30	&  --30$\pm$30  & young	    & $(1)$, $(3)$      \\ 
\object{2MASS J06441439--2841417}&    	&			& 06 44 14.4	& --28 41 42	& +155$\pm$11	&  --36$\pm$11  & M8.0V     & $(1)$	        \\ 
\object{2MASS J06572547--4019134}&    	&			& 06 57 25.5	& --40 19 14	&--220$\pm$30	&   +26$\pm$11  & M7.5V     & $(1)$, $(2)$      \\ 
\object{2MASS J07193188--5051410}&    	&			& 07 19 31.9	& --50 51 41	& +140$\pm$30	&  --10$\pm$30  & L0.0V     & $(1)$	        \\ 
\object{SSSPM J0829--1309}	&    	&			& 08 28 34.2	& --13 09 20	&--490$\pm$40	&   +10$\pm$40  & L2.0V     &		        \\ 
\object{2MASS J08293244--0238543}&      &			& 08 29 32.4	& --02 38 54	&   +4$\pm$6	&   --3$\pm$6	& M8V.0     & $(1)$, $(3)$      \\ 
\object{2MASS J08354256--0819237}&      &			& 08 35 42.6	& --08 19 24	&--730$\pm$180  &  +310$\pm$170 & L5.0V     & $(1)$	        \\ 
\object{2MASS J08472872--1532372}&      &			& 08 47 28.7	& --15 32 37	&--130$\pm$160  &  --20$\pm$160 & L2.0V     & $(1)$, $(3)$      \\ 
\object{2MASS J08500174--1924184}&      &			& 08 50 01.8	& --19 24 18	&--144$\pm$17	&   +49$\pm$17  & M8.0V     & $(1)$	        \\ 
\object{LP 666--9}		&       & GJ 3517		& 08 53 36.2	& --03 29 32	&--156$\pm$9	& --139$\pm$9	& M9.0V     &		        \\ 
\object{2MASSI J0902146--064209}&       &			& 09 02 14.6	& --06 42 10	&  +14$\pm$10	&  --39$\pm$9	& M7.0V     & $(1)$	        \\ 
\object{2MASS J09033514--0637336}&      &			& 09 03 35.1	& --06 37 34	& --73$\pm$7	&   +13$\pm$6	& M7.0V     & $(1)$	        \\ 
\object{DENIS--P J0909.9--0658}	& AB    &			& 09 09 57.5	& --06 58 19	&--280$\pm$190  &  +110$\pm$180 & L0.0V     & $(1)$	        \\ 
\object{2MASS J09130443-0733042}&       &			& 09 13 04.4	& --07 33 04	& --50$\pm$50	& --200$\pm$50  & M9.0V     & $(1)$	        \\ 
\object{SIPS J0921--2104}	&       &			& 09 21 14.1	& --21 04 45	& +100$\pm$60	& --900$\pm$60  & L2.0V     &		        \\ 
\object{2MASS J09263320--0151026}&      &			& 09 26 33.2	& --01 51 03	&--137$\pm$5	&  --31$\pm$5	& M6.0V     & $(1)$	        \\ 
\object{LP 728--52}		&       & NLTT 22091		& 09 34 29.2	& --13 52 43	&--240$\pm$15	& --143$\pm$13  & M7.0:V    &		        \\ 
\object{DENIS J095221.9--192432}&       &			& 09 52 21.9	& --19 24 32	& --68$\pm$5	& --107$\pm$5	& M7.0V     & 			\\ % $(1')$
\object{LP 609--24}		&       & LHS 5165		& 10 03 19.2	& --01 05 08	&--250$\pm$11	&   +32$\pm$9	& M7.0V     &		        \\ 
\object{LP 789--23}		&       & NLTT 23415		& 10 06 32.0	& --16 53 27	&--280$\pm$20	&  +194$\pm$17  & M7.5V     &		        \\ 
\object{2MASS J10184314--1624273}&      &			& 10 18 43.2	& --16 24 27	&  +31$\pm$9	&  --22$\pm$8	& M7.5V     & $(1)$	        \\ 
\object{DENIS--P J102132.3--204407}&    & 			& 10 21 32.3	& --20 44 07	&--339$\pm$12	&  --50$\pm$12  & M8.0V     & $(1)$		\\ 
\object{LP 610--5}		&       & NLTT 24132		& 10 21 51.3	& --03 23 10	& +210$\pm$14	& --151$\pm$10  & M6.5V     &		        \\ 
\object{SDSS J104524.00--014957.6}&     &			& 10 45 24.0	& --01 49 58	&--520$\pm$40	&  --30$\pm$30  & L1.0V     & $(1)$	        \\ 
\object{LP 731--58}		&       & GJ 3622		& 10 48 12.6	& --11 20 08	& +570$\pm$50	&--1500$\pm$60  & M6.5V     &		        \\ 
\object{DENIS--P J104814.7--395606}&    & 			& 10 48 14.6	& --39 56 06   &--1470$\pm$100  & --700$\pm$80  & M8.5V	    & 			\\ % $(1')$
\object{SDSS J104842.81+011158.2}&      & 			& 10 48 42.8	& --01 11 58	&--440$\pm$40	& --240$\pm$30  & L1.0V     & $(1)$		\\ 
\object{DENIS--P J1058.7--1548} &       &			& 10 58 47.9	& --15 48 17	& --60$\pm$160  &  +210$\pm$150 & L3.0V     &		        \\ 
\object{2MASS J11043351--0510439}&      &			& 11 04 33.5	& --05 10 44	&--101$\pm$8	&  --48$\pm$6	& M6.0V     & $(1)$	        \\ 
\object{LP 731--47}		&       & BR B1104--1227  	& 11 06	56.9	& --12 44 02	&--320$\pm$15	& --14$\pm$13	& M6.0V     &		        \\ 
\object{LP 732--20}		&       & LHS 2397		& 11 20 26.4	& --14 40 02	&--367$\pm$16	& --377$\pm$14  & M8.5V     &		        \\ 
\object{2MASS J11304761--2210335}&      &			& 11 30 47.6	& --22 10 34	&--146$\pm$16	& --245$\pm$15  & M8.0V     & $(1)$	        \\ 
\object{LP 673--63}		&       & 			& 11 36 41.0	& --07 55 12	&--190$\pm$20	&  +192$\pm$16  & M6.0V     &		        \\ 
\object{DENIS J114144.0--223215}&    	&			& 11 41 44.0	& --22 32 15	&--190$\pm$20	&  +430$\pm$20  & M8.0V     & 			\\ % $(1')$
\object{2MASS J11553952--3727350}&      &			& 11 55 39.5	& --37 27 35	&  +13$\pm$15	& --778$\pm$13  & L2.0V     & $(2)$ 		\\ % $(1')$
\object{2MASSI J1158027--254536}&       &			& 11 58 02.7	& --25 45 37	& --80$\pm$14	& --167$\pm$12  & M9.0V     & $(1)$	        \\ 
\object{LP 908--5}		&       & NLTT 29333		& 12 01 42.1	& --27 37 46	&--229$\pm$12	&   +19$\pm$11	& M5.5V     &		        \\ 
\object{2MASS J12023666--0604054}&	& 			& 12 02 25.6	& --06 04 05	& +400$\pm$300  &  +300$\pm$200 & M8.0V     & $(1)$, $(3)$      \\ 
\object{2MASS J12022564--0629026}&	& 			& 12 02 36.7	& --06 29 03	& +200$\pm$300  & --100$\pm$200 & M9.0V     & $(1)$, $(3)$      \\ 
\object{LP 734--87}		&       & NLTT 30173		& 12 16 10.1	& --11 26 10	&  +40$\pm$20	& --240$\pm$16	& M5.5V     &		        \\ 
\object{2MASS J12185957--0550282}&      &			& 12 18 59.6	& --05 50 28	&--330$\pm$70	&   +10$\pm$60  & M8.5V     & $(1)$		\\ 
\object{LP 908--68}		&       & LHS 325 a		& 12 23	56.3	& --27 57 47   &--1600$\pm$800	& +800$\pm$800	& M6.0V     &		        \\ 
\object{BRI B1222--1221}	&       &			& 12 24 52.2	& --12 38 35	&--270$\pm$30	& --220$\pm$20  & M9.0V     &		        \\ 
\object{LP 909--55}		&       & 			& 12 36 15.3	& --31 06 46	& +161$\pm$7	&  --78$\pm$8	& M5.5V     &		        \\ 
\object{2MASS J12372705--2117481}&      &			& 12 37 27.0	& --21 17 48	&--222$\pm$9	&  --42$\pm$7	& M6.0V     & $(1)$	        \\ 
\object{2MASS J12473570--1219518}&      &			& 12 47 35.7	& --12 19 52	& --30$\pm$30	& --260$\pm$20  & M8.5V     & $(1)$	        \\ 
\object{Kelu 1}			& AB    &			& 13 05 40.2	& --25 41 06	&--310$\pm$12	&  --12$\pm$10  & L2.0V+... &		        \\ 
\object{CE 303}			&       &			& 13 09 21.8	& --23 30 35	&  +17$\pm$9	& --372$\pm$8	& M8.0V     &		        \\ 
\object{2MASS J13300232--0453202}&      &			& 13 30 02.3	& --04 53 20	& --79$\pm$9	&   --7$\pm$9	& M8.0V     & $(1)$	        \\ 
\object{2MASS J13322442--0441126}&      &			& 13 32 24.4	& --04 41 13	&  +59$\pm$19	&  --10$\pm$16  & M7.5V     & $(1)$	        \\ 
\object{2MASS J13392651--1755053}&      &			& 13 39 26.5	& --17 55 05	&--190$\pm$20	&  --70$\pm$20  & M7.5V     & $(1)$	        \\ 
\object{2MASS J13401152--1451591}&      &			& 13 40 11.5	& --14 51 59	&--101$\pm$17	& --190$\pm$12  & M6.5V     & $(1)$	        \\ 
\object{LP 911--56}		&       & CE 359		& 13 46	46.1	& --31 49 26  	&--332$\pm$18	& +154$\pm$17	& M6.0V     & $(2)$		\\ 
\object{DENIS--P J135714.9--143852}&    &			& 13 57	15.0	& --14 38 53 	& --38$\pm$10	& --106$\pm$9   & M7.5V     & 	        	\\ 
\object{DENIS--P J141121.2--211950}&    &			& 14 11 21.3	& --21 19 50	& --58$\pm$9	& --102$\pm$8	& M9.0V     & $(1)$	        \\ 
\object{2MASS J14135981--0457483}&      &			& 14 13 59.8	& --04 57 05	&--190$\pm$50	&  --60$\pm$40  & M8.0V     & $(1)$	        \\ 
\object{2MASS J14211873--1618201}&      &			& 14 21 18.7	& --16 18 20	&--230$\pm$20	&  --70$\pm$20  & M7.5V     & $(1)$	        \\ 
\object{2MASS J14241870--3514325}&      &			& 14 24 18.7	& --35 14 32	&  --2$\pm$7	&  --79$\pm$6	& M6.5V     & $(1)$	        \\ 
\object{Proxima Centauri}	&       & $\alpha$ Cen C	& 14 26 19.0	& --62 28 04   &--3775$\pm$2    &+769.3$\pm$1.3 & M5.5V	    & $(4)$	        \\ 
\object{G 124--62}		& A  	& 			& 14 41 35.8	& --09 46 39	&--208$\pm$9	&  --26$\pm$9   & M4.5Ve    &		        \\ 
				& BC  	& DE1441--27AB		& 14 41 37.2	& --09 45 59	&--190$\pm$80	&   +60$\pm$80  & L1.0V+... &		        \\ 
\object{DENIS J145601.3--274736}&    	&			& 14 56 01.4	& --27 47 37	&--180$\pm$18	& --204$\pm$16  & M9.0V     & 			\\ % $(1')$
\object{LP 914--54}		&       & GJ 3877		& 14 56 38.3	& --28 09 47	&--470$\pm$40	& --900$\pm$60  & M7.0V     &		        \\ 
\object{TVLM 868--54745}	&       &			& 15 00 34.3	& --00 59 45	&  +82$\pm$6	&   --6$\pm$6	& M8.0:V    & $(1)$	        \\ 
\object{2MASS J15072779--2000431}&      &			& 15 07 27.8	& --20 00 43	& +109$\pm$9	&  --78$\pm$9	& M7.5V     & $(1)$	        \\ 
\object{DENIS--P J151233.3--103241}&    &			& 15 12 33.3	& --10 32 41	& --40$\pm$20	&  --37$\pm$19  & M8.5V     & $(1)$, $(3)$      \\ 
\object{2MASS J15551573--0956055}&      &			& 15 55	15.7	& --09 56 06	&--400$\pm$1200	&--1900$\pm$1100& L1.0V     & $(3)$	        \\ 
\object{LSR J1610--0040}	&       & 			& 16 10	29.0	& --00 40 53	&--680$\pm$90	&--1250$\pm$90	& sdM/L:    &		        \\ 
\object{LP 624--54}		&       & 			& 16 14 25.2	& --02 51 01	&   +6$\pm$14	&  +350$\pm$14	& M6.0V     &		        \\ 
\object{2MASS J16452211--1319516}&      &			& 16 45 22.1	& --13 19 52	&--360$\pm$40	& --820$\pm$40  & L1.5V     & $(1)$	        \\ 
\object{LP 626--2}		&       & 			& 16 45	28.2	& --01 12 29	&  --8$\pm$12	& --226$\pm$13	& M5.5V     &		        \\ 
\object{V1054 Oph}          	& A--BE	& GJ 664AB		& 16 55 28.8	& --08 20 10	&--900$\pm$50	& --910$\pm$50  & M3.0Ve+...& $(2)$	        \\ % -829$\pm$4 --879$\pm$2 Perryman et al. (1997)    
			        & C	& GJ 663		& 16 55 25.3	& --08 19 21	&--830$\pm$30	& --920$\pm$30  & M4.0V+... &		        \\ % -813$\pm$4 --895$\pm$2 Perryman et al. (1997)    
				& D	& vB 8			& 16 55 35.3	& --08 23 40	&--790$\pm$20	& --900$\pm$20  & M7.0V	    & $(2)$	        \\ % -771 --871 Deacon et al. (2005) 
\object{SCR J1845--6357}	& AB    & 			& 18 45 05.4	& --63 57 48	&+2440$\pm$100  &  +700$\pm$120 & M8.5V+... &		        \\ 
\object{2DENIS--P J200213.4--542555}&   &			& 20 02	13.4	& --54 25 56	&  +49$\pm$7	& --367$\pm$8 	& M6.0V     & $(2)$		\\       
\object{2MASSI J2004536--141622}&       &			& 20 04 53.7	& --14 16 23	& +534$\pm$18	&   +56$\pm$17  & M7.5V     & $(1)$	        \\ % red      
\object{2MASS J20140359--2016217}&      &			& 20 14 03.6	& --20 16 22	&  +25$\pm$11	& --124$\pm$12  & M7.5V     & $(1)$	        \\ 
\object{2MASS J20151945--1601334}&      &			& 20 15 19.4	& --16 01 34	& --34$\pm$5	& --101$\pm$5	& M5.5V     & $(1)$	        \\ 
\object{2MASS J20192695--2502441}&      &			& 20 19 27.0	& --25 02 44	&--130$\pm$20	&  --90$\pm$20  & M8.0V     & $(1)$	        \\ 
\object{2MASS J20335733--0429413}&      &			& 20 33 57.3	& --04 29 41	&  +33$\pm$13	& --257$\pm$13  & M6.5V     & $(1)$	        \\ 
\object{2MASS J20370715--1137569}& AB   &			& 20 37 07.2	& --11 37 57	& --30$\pm$30	& --390$\pm$30  & M8.0V+... & $(1)$	        \\ 
\object{2MASS J20391314--1126531}&      &			& 20 39 13.1	& --11 26 53	&  +64$\pm$11	& --105$\pm$11  & M8.0V     & $(1)$	        \\ 
\object{LP 695-351}		&       &			& 20 41 41.0	& --03 33 53	& +166$\pm$5	&  --67$\pm$5	& M6.0V     &		        \\ 
\object{2MASS J20473176--0808201}&	& 			& 20 47 31.8	& --08 08 20	& +100$\pm$300  & --200$\pm$300 & M7.0V     & $(1)$, $(3)$      \\ 
\object{2MASS J20491972--1944324}&      &			& 20 49 19.7	& --19 44 32	& +179$\pm$9	& --279$\pm$9	& M7.5V     & $(1)$	        \\ 
\object{DENIS--P J205754.1--025229}&    &			& 20 57 54.1	& --02 52 30	&  +20$\pm$30	&  --90$\pm$30  & L1.5V     & $(1)$	        \\ 
\object{2MASS J21041491--1037369}&      &			& 21 04 14.9	& --10 37 37	& +550$\pm$170  & --180$\pm$170 & L3.0V     & $(1)$	        \\ 
\object{2MASS J21130293--1009412}& AB   &			& 21 13 02.9	& --10 09 41	& --24$\pm$6	& --122$\pm$7	& M6.0V+... & $(1)$	        \\ 
\object{2MASS J21254581--0018340}&      &			& 21 25 45.8	& --00 18 34	&  +16$\pm$6	&   +22$\pm$7	& M6.5V     & $(1)$, $(3)$      \\ 
\object{LP 698--2}		&       & NLTT 51488		& 21 32 29.8	& --05 11 58	& +126$\pm$11	& --352$\pm$12  & M6.0V     &		        \\ 
\object{LP 759--17}		&       &			& 22 02 11.3	& --11 09 46	& +133$\pm$11	& --192$\pm$11  & M6.5V     &		        \\ 
\object{LP 759--25}		&       & NLTT 52882		& 22 05	35.8	& --11 04 29	&--173$\pm$8	& --104$\pm$8   & M5.5V     &		        \\ 
\object{DENIS J220622.7--204706}& AB 	&			& 22 06 22.8	& --20 47 06	&  +30$\pm$9	&  --41$\pm$10  & M8.0V+... &		        \\ 
\object{2MASSI J2214506--131959}&       &			& 22 14 50.7	& --13 19 59	& +255$\pm$15	& --263$\pm$15  & M7.5V     & $(1)$	        \\ 
\object{2MASS J22263689--0239502}&      &			& 22 26 36.9	& --02 39 50	& +210$\pm$19	&  --81$\pm$19  & M6.5V     & $(1)$	        \\ 
\object{EZ Aqr}			& ABC   &			& 22 38 33.6	& --15 17 59	&+2214$\pm$5	& +2295$\pm$5	& M5.5VST   & $(4)$	        \\ 
\object{LP 700--66}		&       & NLTT 54525		& 22 40 38.6	& --02 50 56	& +197$\pm$13	& --241$\pm$13  & M6.5V     &		        \\ 
\object{2MASSI J2252014--181558}&       &			& 22 52 01.5	& --18 16 00	&  +81$\pm$15	& --382$\pm$16  & M8.5V     & $(1)$	        \\ 
\object{DENIS--P J225451.8--284025}&    &			& 22 54 51.9	& --28 40 25	& --50$\pm$50	&   +60$\pm$50  & L0.5V     & $(1)$, $(3)$      \\ 
\object{2MASS J23062928--0502285}&      &			& 23 06 29.3	& --05 02 28	& --76$\pm$7	&  --95$\pm$7	& M7.5V     & $(2)$	        \\ % +905$\pm$ --516$\pm$ Deacon et al. (2005)
\object{LP 702--58}		&       & NLTT 56373		& 23 17 20.7	& --02 32 32	& +213$\pm$15	&  --92$\pm$17  & M6.5V     &		        \\ 
\object{HD 221356}		&    	& BD--04 5896		& 23 31 31.5	& --04 05 14	& +178.6$\pm$1.0&--192.8$\pm$0.8& F8.0V     & $(4)$	        \\ 
				& AB   	& 2M2331--04AB		& 23 31 01.6	& --04 06 19	& +220$\pm$20	& --190$\pm$20  & M8.0V+... &		        \\ 
\object{LP 732--20}		&       &			& 23 37 14.9	& --08 38 08	& +248$\pm$13	&   +19$\pm$13  & M7.0V     &		        \\ 
\object{LP 763--3}		&       & NLTT 57439		& 23 37 38.3	& --12 50 28	& +218$\pm$13	& --317$\pm$12  & M6.0V     &		        \\ 
\object{LEHPM 6334}		&    	&  			& 23 51	50.4	& --25 37 37	& +376$\pm$18	&  +158$\pm$18  & M9.0V     & $(2)$      	\\ 
\object{DENIS--P J235359.4--083331}&    &			& 23 53 59.5	& --08 33 31	&--580$\pm$200	&  --20$\pm$200 & M8.5V     &       		\\ 
\object{LEHPM 6494}		&    	& SSSPM J2356--3426 	& 23 56	10.8	& --34 26 04	&  +90$\pm$13	& --306$\pm$13  & M9.0V     &       		\\ 
\enddata
\tablenotetext{a}{Remarks -- 
		$(1)$: first proper motion measurement;
		$(2)$: double detection in SSA;
		$(3)$: high $\delta \mu / \mu$ ratio;
		$(4)$: proper motion from the literature.}
\end{deluxetable}

\clearpage

%of \object{BD--02 912}. 

\clearpage

\begin{figure}
%\epsscale{0.5}
%\plottwo{f2a.ps}{f2b.ps}
\caption{False-color composite images centered on the system LP 655--23 +
2M0430--08 (Koenigstuhl~2~AB). 
Red, green and blue are for photograhic $I_N$ (UKST), $R$ (POSS--I) and $B_J$
(UKST), taken at epochs separated by 46.0 years.
North is up, east is left.
{\em Left window:} 10 $\times$ 10\,arcmin$^2$ field of view. 
{\em Right window:} zoom of the left window, 2.5 $\times$ 2.5\,arcmin$^2$ field
of view. 
{\em AVAILABLE only in ApJ}. 
\label{rgb2}}
\end{figure}

\clearpage

\begin{figure}
\epsscale{0.5}
%\plotone{f3.ps}
\caption{Same as left window in Fig.~\ref{rgb2}, but for the system HD~221356 +
2M2331--04AB (Koenigstuhl~3~A--BC). 
Epochs are separated by 48.3 years.
{\em AVAILABLE only in ApJ}. 
\label{rgb3}}
\end{figure}

\end{document}